\documentclass{aa}
\usepackage{epsfig}
\usepackage{rotating}
\usepackage{natbib}
\usepackage{graphicx}

\newcommand{\chan}{{\sl Chandra}}
\newcommand{\rosat}{{\sl ROSAT}}
\newcommand{\asca}{{\sl ASCA}}

\newcommand{\xmm}{{\sl XMM}}
\newcommand{\rxte}{{\sl RXTE}}

\newcommand{\vlt}{{\sl VLT}}

\newcommand{\ntt}{{\sl NTT}}
\newcommand{\susi}{{\sl SUSI2}}
\newcommand{\nttn}{{\sl New Technology Telescope}}

\newcommand{\naco}{{\sl NACO}}
\newcommand{\nacon}{{\sl NAos COnica}}

\newcommand{\tmass}{{\sl 2MASS}}
\newcommand{\gsc}{{\sl GSC-2}}

\newcommand{\dao}{{\tt Daophot}}

\begin{document}

 \title{Optical and Infrared Observations of the X-ray source 1WGA J1713.4$-$3949 in the G347.3-0.5 SNR\thanks{Based on observations collected at the European Southern Observatory, Paranal, Chile under  programme ID 073.D-0632(A),077.D-0764(A) } }

 \author{R. P. Mignani\inst{1}
 \and
 S. Zaggia\inst{2}
 \and 	
 A. De Luca\inst{3}
\and
R. Perna\inst{4}
\and 	
 N. Bassan \inst{3}
\and 	
 P. A. Caraveo\inst{3}
}

   \offprints{R. P. Mignani}

   \institute{Mullard Space Science Laboratory, University College London, Holmbury St. Mary, Dorking, Surrey, RH5 6NT, UK\\
              \email{rm2@mssl.ucl.ac.uk}
\and
INAF, Osservatorio Astronomico di Padova, Vicolo dell'Osservatorio 5, Padua, 35122, Italy \\
 \email{simone.zaggia@oapd.inaf.it}
\and
INAF, Istituto di Astrofisica Spaziale, Via Bassini 15, Milan, 20133, Italy \\
 \email{[deluca,bassan,pat]@iasf-milano.inaf.it}
\and
JILA and Department of Astrophysical and Planetary Sciences, University of Colorado, 440 UCB, Boulder, 80309, USA
\email{rosalba@jilau1.Colorado.EDU}
}

     \date{Received ...; accepted ...}

   \abstract {X-ray  observations unveiled the  existence of enigmatic
point-like sources  at the centre of young  supernova remnants (SNRs).
These sources,  dubbed Central Compact Objects (CCOs),  are thought to
be neutron  stars formed by  the supernova explosion.   However, their
multi-wavelength phenomenology is  surprisingly different from that of
most young neutron  stars.}{The aim of this work  is to understand the
nature of the  CCO 1WGA J1713.4$-$3949 in the  G347.3-0.5 SNR, through
deep optical  and IR observations,  the first ever performed  for this
source.}{By exploiting  its derived  \chan\ X-ray position  we carried
out optical (BVI)  observations with the \ntt\ and  Adaptive Optics IR
(JHK$_s$)  observations  with  the  \vlt.   }{We  detected  two  faint
(I$\approx23.5$, I$\approx 24.3$.) patchy objects in the \ntt\ images,
close to the  \chan\ error circle.  They were  clearly resolved in our
\vlt\  images which  unveiled a  total of  six  candidate counterparts
($17.8<H<20.3$)  with quite red  colours (H-K$_s$$\sim$0.6).   If they
are stars, none of them can be associated with 1WGA J1713.4$-$3949 for
the most likely  values of distance and hydrogen  column density.  The
identification of the faintest  candidate with the neutron star itself
can not be  firmly excluded, while the identification  with a fallback
disk  is ruled  out by  its  non-detection in  the I  band.  No  other
candidates are detected down to B$\sim 26$, V$\sim26.2$, I$\sim 24.7$,
H$\sim$21.3  and  K$\sim$20.5.  }{Our  high-resolution  IR imaging  of
unveiled a few objects close/within  the \chan\ X-ray position of 1WGA
J1713.4$-$3949.   However,  at present  none  of  them  can be  firmly
identified as its likely counterpart. }


\titlerunning{Optical and infrared observations of 1WGA J1713.4$-$3949}

   \maketitle

\section{Introduction} 

X-ray observations have  unveiled the existence
of peculiar classes of Isolated Neutron Stars (INSs) which stand apart
from the family of more  classical radio pulsars in being radio-silent
and  not powered  by the  neutron star  rotation but  by  still poorly
understood emission mechanisms.  Some  of the most puzzling classes of
radio-silent INSs are identified with a group of X-ray sources detected
at the centre of young ($\sim$ 10-40 kyears) supernova remnants (SNRs),
hence dubbed  Central Compact  Objects  or CCOs  (Pavlov et  al.
2002).

Although the SNR associations imply ages of the order of a few kyears,
their X-ray  properties make CCOs completely different  from the other
young INSs  in SNRs (Pavlov  et al. 2004;  De Luca 2008).  Oly  two of
them exhibit X-ray  pulsations, with periods in the  $\sim$ 100-400 ms
range, and the  measured upper limits on the  period derivatives yield
spin down ages  $\ge 10^3$ exceeding the SNR  age.  Furthermore, their
X-ray spectra  are not purely  magnetospheric but have  strong thermal
components.   Finally, they are  not embedded  in pulsar  wind nebulae
(PWN).  The discovery of long  term X-ray flux variations (Gotthelf et
al.  1999) and of a 6.7  hours periodicity (e.g. De Luca et al.  2006)
in the RCW 103 CCO  further complicated the picture, suggesting either
a binary system  with a low-mass companion, or  a long-period magnetar
(De Luca et  al.  2006; Pizzolato et al.  2008).   For other CCOs, the
invoked  scenarios involve  low-magnetised INSs  surrounded  by debris
disks  formed after  the supernova  event (Gotthelf  \&  Halpern 2007;
Halpern  et al.   2007),  isolated accreting  black  holes (Pavlov  et
al.  2000),  and dormant  magnetars  (Krause  et  al. 2005).   In  the
optical/IR, deep  observations have been performed only  for a handful
of objects  (see De Luca  2008 for a  summary) but no  counterpart has
been identified yet,  with the possible exception of  the Vela Jr. CCO
(Mignani et al. 2007a).

One of  the CCOs which still  lack a deep  optical/IR investigation is
1WGA  J1713.4$-$3949 in the  young ($\le  40$ kyears)  G347.3-0.5 SNR.
Discovered by \rosat\ (Pfeffermann \& Aschenbach 1996), the source was
re-observed  with  \asca\  and  identified   as  an  INS  due  to  its
high-temperature  spectrum  and the  lack  of  an optical  counterpart
(Slane  et al.   1999). 1WGA  J1713.4$-$3949 was  later  observed with
\rxte, \chan\ and \xmm\ (Lazendic et al.  2003; Cassam-Chena\"i et al.
2004),  with  all  the  observation  consistent with  a  steady  X-ray
emission. The  X-ray luminosity is  $L_{0.5-10 keV} \sim 6  ~ 10^{34}$
(d/6 kpc)$^2$  erg s$^{-1}$, where  6 kpc is the  originally estimated
SNR distance  (Slane et  al.  1999).  A  revised distance of  $1.3 \pm
0.4$ kpc was recently obtained by Cassam-Chena\"i et al.  (2004).  The
X-ray spectrum  can be fitted  either by a blackbody,  likely produced
from hot polar caps, plus  a power-law ($kT\sim$ 0.4 keV; $\Gamma \sim
4$;  $N_H  \sim  9  ~  10^{21}$  cm$^{-2}$),  or  by  two  blackbodies
($kT_1\sim$  0.5 keV;  $kT_2\sim$  0.3  keV; $N_H  \sim  5 ~  10^{21}$
cm$^{-2}$).  No  X-ray pulsations were  detected so far (Slane  et al.
1999; Lazendic et  al.  2003), nor any radio  counterpart (Lazendic et
al.  2004), thus strengthening the  case for 1WGA J1713.4$-$3949 to be
a member of the CCO class.

Here we present the results of the first, deep optical/IR observations
of   1WGA   J1713.4$-$3949   performed   with  the   ESO   telescopes.
Observations are described in Sect. 2, while the results are described
and discussed in Sect 3 and 4, respectively.

\section{Observations and data reduction}

\subsection{Optical observations}

1WGA J1713.4$-$3949  was observed  on June
13th 2004 at the ESO La  Silla Observatory with the \nttn\ (\ntt). The
telescope was equipped  with the second generation of  the {\em SUperb
Seeing   Imager}   (\susi).   The   camera   is   a   mosaic  of   two
2000$\times$4000 pixels  EEV CCDs  with a 2$\times$2 binned
pixel  scale  of 0\farcs16  (5\arcmin.5  $\times$  5\arcmin.5 field  of
view).

Repeated  exposures  were obtained  in  the  broad-band  B, V,  and  I
filters.  The \susi\ observations log  is summarised in the first half
of Table \ref{data}. Observations were performed with the target close
to  the zenith  and  under reasonably  good  seeing conditions  ($\sim
1\arcsec$).  Since the target was  always centred on the left chip, no
dithering was applied  to the B and V-band  exposures while the I-band
ones were  dithered to compensate  for the fringing  pattern affecting
the CCD at longer wavelengths.   Both night (twilight flat fields) and
day time  calibration frames (bias,  dome flat fields)  were acquired.
Unfortunately, due to the presence of clouds both at the beginning and
at the end  of the night no standard  star observations were acquired.
As  a reference  for  the  photometric calibration  we  then used  the
closest  in time zero  points regularly  measured using  Landolt stars
(Landolt  1983)  as  part  of  the  instrument  calibration  plan  and
available  in the  photometry calibration  database maintained  by the
\ntt/\susi\ team.  According to the zero point trending plots
\footnote{http://www.ls.eso.org/lasilla/sciops/ntt/susi/docs/susiCounts.html},
we estimate a conservative uncertainty of $\sim 0.1$ magnitudes on the
values extrapolated to the night of our observations.

\subsection{Infrared observations}

1WGA J1713.4$-$3949 was observed on
May 23rd and  24th 2006 at the ESO  Paranal Observatory with \nacon\
(\naco), the  adaptive optics (AO) imager and  spectrometer mounted at
the \vlt\ Yepun  telescope.  In order to provide  the best combination
between  angular resolution and  sensitivity, we  used the  S27 camera
with a pixel scale of  0\farcs027 ($28''\times28''$ field of view).
As  a  reference  for  the  AO  correction  we  used  the  \gsc\  star
S230012111058 ($V=14.3$), positioned 11\farcs5 away from our target, with
the $VIS$ ($4500-10000
~ $ \AA) dichroic  element and wavefront sensor.   

Observations were  performed in the H  and K$_s$ bands.   To allow for
subtraction of  the variable IR  sky background, each  integration was
split in sequences of  short randomly dithered exposures with Detector
Integration Times (DIT) of 24 s and 5 exposures (NDIT) along each node
of the  dithering pattern.  The \naco\ observations  log is summarised
in the  second half  of Table \ref{data}.   For all  observations, the
seeing   conditions   were   on   average   below   $\sim   0\farcs8$.
Unfortunately, since the target was  always observed at the end of the
night,  the  airmass  was   always  above  1.4.  Sky  conditions  were
photometric in both nights.  On  the first night, the second and third
K$_s$-band  exposure  sequence  were  aborted because  the  very  high
airmass prevented Because of their  worse image quality and their much
lower  signal--to--noise,  these  data   are  not  considered  in  the
following analysis.  The K$_s$-band  exposure sequence obtained on the
second night was  interrupted despite of the very  good seeing because
of the  incoming twilight.  Thanks  to the combination of  good seeing
and low  airmass, the H-band exposure  is the one with  the best image
quality.  Night (twilight flat fields) and day time calibration frames
(darks,  lamp flat  fields) were  taken daily  as part  of  the \naco\
calibration  plan.  Standard  stars from  the Persson  et  al.  (1998)
fields were observed in both nights for photometric calibration.

\begin{table}
\begin{center}
  \caption{Log of  the \ntt/\susi\ (first half)  and \vlt/\naco\ (second half) observations of  the 1WGA J1713.4$-$3949 field. Columns report the observing epoch, the filter, the total integration time, and the average seeing and airmass. }
\begin{tabular}{ccccc} \\ \hline
 yyyy-mm-dd  & Filter & T (s) & Seeing (``) & Airmass	\\ \hline
 2004-06-13  &      B    &   3200    &   1.14  &   1.07  \\
 2004-06-13  &      V    &   6400    &   1.12  &   1.03  \\
 2004-06-13  &      I    &   3150    &   1.0  &   1.16  \\ \hline
 2006-05-24  &      Ks   &   1800    &   0.66  &   1.45  \\
 2006-05-24  &      Ks   &    360    &   0.95  &   1.70  \\
 2006-05-24  &      Ks   &    600    &   0.78  &   1.81  \\
 2006-05-25  &      H    &   2400    &   0.62  &   1.42  \\
 2006-05-25  &      Ks   &   1200    &   0.40  &   1.74  \\ \hline
\end{tabular}
\label{data}
\end{center}
\end{table}

\subsection{Data reduction}

The \ntt/\susi\ data were  reduced using standard routine available in
the           {\tt           MIDAS}           data           reduction
package\footnote{http://www.eso.org/sci/data-processing/software/esomidas/}.
After the  basic reduction steps  (hot pixels masking, removal  of bad
CCD column,  bias subtraction, flat field  correction), single science
frames  were combined  to filter  cosmic ray  hits and  to  remove the
fringing patterns  in the I-band.   The astrometry was  computed using
the coordinates  and positions of  61 stars selected from  the \tmass\
catalogue (Skrutskie et  al. 2006).  For a better  comparison with the
\vlt/\naco\ IR images, the I-band  image was taken as a reference. The
pixel coordinates of  the \tmass\ stars (all non  saturated and evenly
distributed  in the field)  were measured  by fitting  their intensity
profiles with a Gaussian function using the {\tt GAIA} ({\tt Graphical
Astronomy               and              Image              Analysis})
tool\footnote{star-www.dur.ac.uk/~pdraper/gaia/gaia.html}.  The fit to
celestial coordinates  was computed  using the {\tt  Starlink} package
{\tt ASTROM}\footnote{http://star-www.rl.ac.uk/Software/software.htm}.
The rms of  the astrometric fit residuals was  $\approx$ 0\farcs09 per
coordinate.   After  accounting for  the  0\farcs2 {\bf  conservative}
astrometric accuracy  of \tmass\ (Skrutskie et al.  2006), the overall
uncertainty to be attached to our astrometry is finally 0\farcs24.

The \vlt\  data were processed  through the ESO \naco\  data reduction
pipeline\footnote{www.eso.org/observing/dfo/quality/NACO/pipeline}.
For each  band, science frames  were reduced with the  produced master
dark and  flat field frames and  combined to correct  for the exposure
dithering and  to produce  cosmic-ray free and  sky-subtracted images.
The photometric calibration was  applied using the zero point provided
by the  \naco\ pipeline,  computed through fixed  aperture photometry.
The  astrometric calibration  was performed  using the  same procedure
described  above.    However,  since  only  five   \tmass\  stars  are
identified in the narrow \naco\  S27 camera field of view, we computed
the astrometric  solution using as a  reference a set  of 23 secondary
stars found in  common with the \susi\ I-band  image, calibrated using
\tmass.  The rms  of the astrometric fit residuals  was then $\approx$
0\farcs06  per coordinate.   By adding  in quadrature  the rms  of the
astrometric fit residuals  of the \susi\ I-band image  and the average
astrometric  accuracy of  \tmass\,  we  thus end  up  with an  overall
accuracy of 0\farcs25 on the \naco\ image astrometry.

\begin{figure*}
\centering 
\includegraphics[bb=5 94 288 384,height=8.2cm,angle=-90,clip]{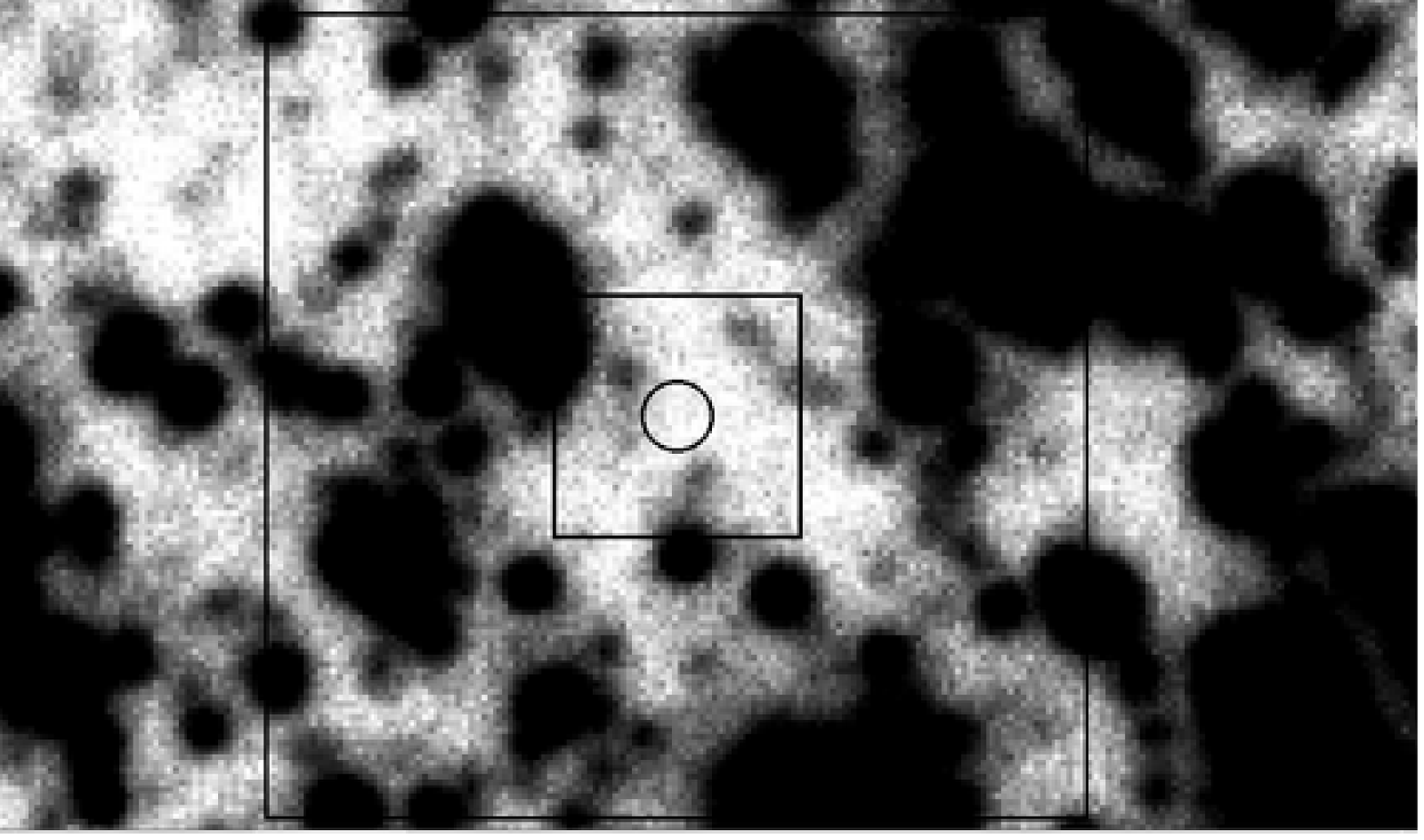}
 \includegraphics[bb=4 115 288 400,width=8cm,angle=-90,clip=]{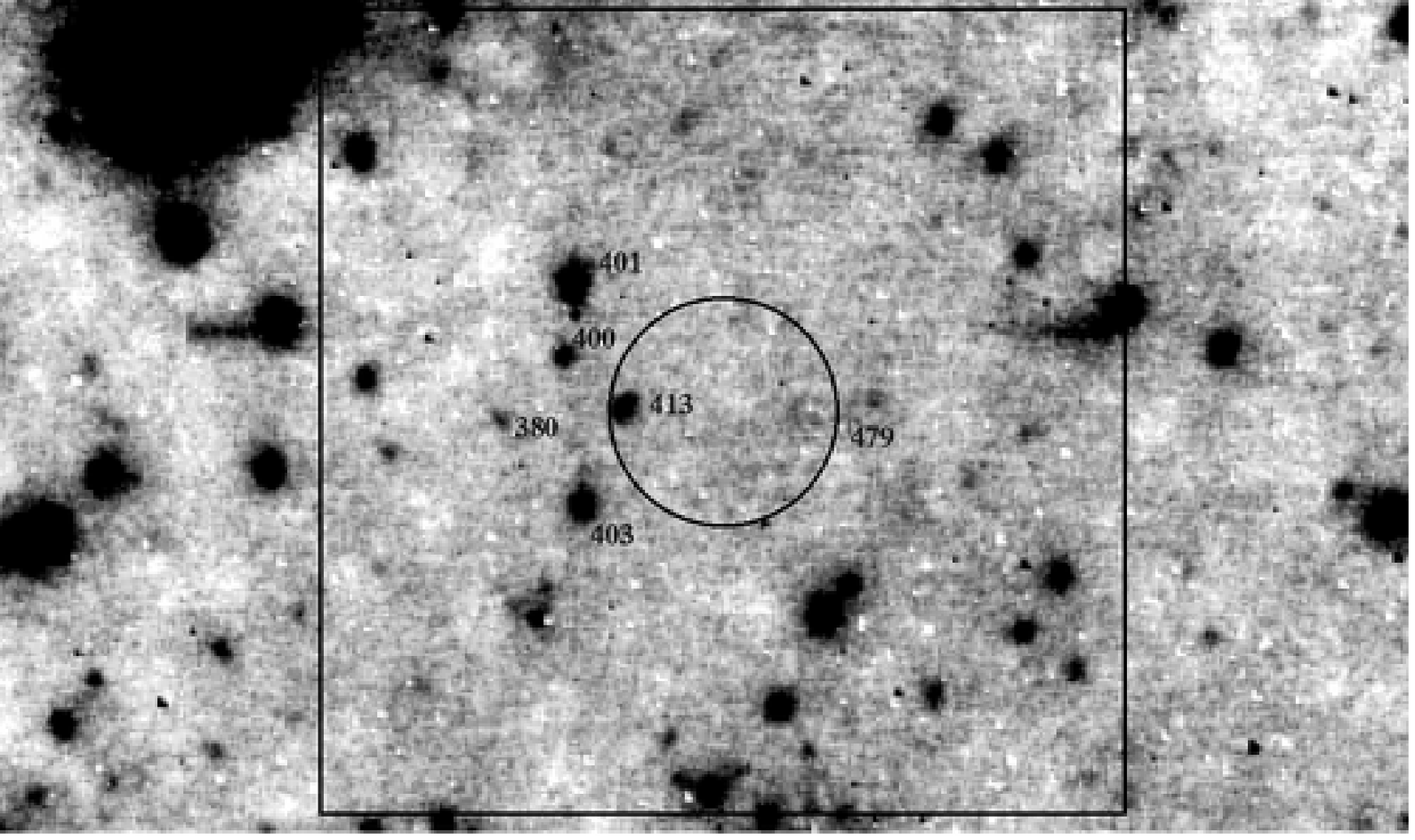}
  \caption{(left) 20\arcsec$\times$20\arcsec \ntt$/$\susi\ I-band  of the 1WGA
J1713.4$-$3949 field. (right) 6\arcsec$\times$6\arcsec \vlt$/$\naco\ H-band image of the same field. The area corresponds to the square overplotted on the left hand side image. The circle  marks the \chan\  position of the X-ray source, where the radius (0\farcs85) accounts both for the intrinsic absolute accuracy of the \chan\ coordinates  (99\% confidence level) and for  the uncertainties of our astrometric calibration (see Sect. 2.3).  Possible counterparts are labelled on the right hand side image, with object 479 detected only at $<5 \sigma$. }
\label{rxj1713_fc}       % Give a unique label
\end{figure*}

\section{Data analysis and results}

\subsection{Astrometry}

We derived the coordinates of 1WGA J1713.4$-$3949 through the analysis
of unpublished  \chan\ observations.   The field of  1WGA J1713.4-3949
was observed on  April 19th 2005 with the  {\sl ACIS/I} instrument for
9.7  ks.  Calibrated  (level 2)  data were  retrieved from  the \chan\
X-ray  Center  Archive  and  were  analysed  using  the  {\tt  Chandra
Interactive Analysis of Observations}  software ({\tt CIAO v3.3}).  In
order to compute the target  position, we performed a source detection
in the  0.5-10 keV energy range  using the {\tt  wavdetect} task.  The
source  coordinates  turned  out   to  be  $\alpha  (J2000)=17^h  13^m
28.32^s$,  $\delta (J2000)=  -39^\circ 49\arcmin  53\farcs34$,  with a
nominal    uncertainty   of    $\sim$   0\farcs8    (99\%   confidence
level)\footnote{http://cxc.harvard.edu/cal/ASPECT/celmon/}.         The
identification of a field X-ray  source with the bright star HD 322941
at a  position consistent with the  one listed in  the Tycho Reference
Catalog (H{\o}g  et al.  2000) confirmed the  accuracy of  the nominal
\chan\ astrometric solution. Unfortunately, since no other field X-ray
source could  be unambiguously identified with  catalogued objects, it
was not  possible to  perform any boresight  correction to  the \chan\
data in order to improve the nominal astrometric accuracy.

The  computed  1WGA  J1713.4$-$3949  position  is shown  in  Fig.   1,
overplotted on  the \ntt/\susi\ I-band  and on the  \vlt/\naco\ H-band
images.   In the  former (Fig.1-left)  a  faint and  patchy object  is
clearly  detected northeast of  the \chan\  error circle  (I$=23.5 \pm
0.3)$ and a fainter one (I$=24.3\pm0.4$) is possibly detected south of
it.  However, in both cases  their patchy structure makes it difficult
to  determine whether  their are  single, or  blended  with unresolved
field objects.   No other  object is detected  within or close  to the
\chan\ error circle down to B$\sim$26, V$\sim$26.2 and I$\sim$24.7 ($3
\sigma$).   However, due to  the better  seeing conditions  (see Table
\ref{data}) and  to the sharper  angular resolution, five  objects are
clearly detected in the  \vlt/\naco\ image (Fig.  1-right).  Of these,
object  413 falls  within the  \chan\ error  circle.  A  sixth fainter
object   (479)   is   possibly    detected,   albeit   at   very   low
significance. They are all  point-like and compatible with the on-axis
\naco\ PSF.   Objects 401  and 403 are  identified with the  two faint
objects  detected  in  the  \ntt/\susi\  I-band  image  northeast  and
southeast of the \chan\  error circle, respectively.  The former might
be actually a  blend of objects 401 and  400, whose angular separation
($\approx 0\farcs6$) is smaller than the PSF of the \ntt/\susi\ image.
We thus take  the measured magnitude (I$=23.5 \pm  0.3$) of object 401
with caution.  All the objects detected in the \naco\ H-band image are
also  detected in the  longest 1200  and 1800  s K$_s$-band  ones (see
Table  \ref{data}).No other  object is  detected close  to  the \chan\
error circle down to H$\sim$21.3 and K$_s\sim$20.5 ($3 \sigma$).

\begin{table}
\begin{center}
  \caption{\vlt/\naco\  H and K$_s$-band photometry and colour of the candidate counterparts  of  1WGA
J1713.4$-$3949.}
\begin{tabular}{cccc} \\ \hline
ID  & H                & K$_s$            & H$-$K$_s$	\\ \hline
380 & 19.75  $\pm$  0.14 & 19.32 $\pm$ 0.11  & 0.43 $\pm$  0.18 \\
400 & 18.98  $\pm$  0.13 & 18.49 $\pm$ 0.09  & 0.49 $\pm$  0.16 \\
401 & 17.82  $\pm$  0.13 & 17.32 $\pm$ 0.08  & 0.50 $\pm$  0.15 \\
403 & 18.47  $\pm$  0.13 & 17.87 $\pm$ 0.08  & 0.60 $\pm$  0.15 \\
413 & 18.63  $\pm$  0.13 & 18.31 $\pm$ 0.09  & 0.33 $\pm$  0.16 \\
479 & 20.34  $\pm$  0.15 & 19.61 $\pm$ 0.12  & 0.73 $\pm$  0.20 \\ \hline
\end{tabular}
\label{phot}
\end{center}
\end{table}

\subsection{Photometry}

We  computed  objects magnitudes  in  the  \naco\  images through  PSF
photometry using the  suite of tools \dao (Stetson  1992) and applying
the same procedures  described in Zaggia et al.  (1997) and applied in
Mignani et al.   (2007a) and De Luca et al.  (2008).  Since the \naco\
PSF is largely oversampled, we re-sampled the images with a $3\times3$
pixels        window        using        the        {\tt        swarp}
program\footnote{http://terapix.iap.fr/}      to      increase     the
signal--to--noise ratio.   As a reference  for our photometry  we used
the co-added  and re-sampled H-band image  to create a  master list of
objects, which we registered on the  K$_s$-band one and used as a mask
for the object detection. For each image, the model PSF was calculated
by  fitting  the profile  of  a  number  of bright  but  non-saturated
reference stars in the field and used to measure the objects fluxes at
the reference positions.  Our photometry was calibrated using the zero
points  provided by the  \naco\ pipeline  after applying  the aperture
correction, with  an attached  errors of $\sim  0.13$ and  $\sim 0.08$
magnitudes in H and  K$_s$, respectively.  Single band catalogues were
then matched and used as a  reference for our colour analysis.  The IR
magnitudes of our  candidates are listed in Tab.  \ref{phot}.  We used
the K$_s$-band  photometry performed on the two  consecutive nights to
search for variability on time  scales of hours.  However, none of our
candidates shows flux variations  larger than 0.1 magnitudes, which is
consistent with  our photometric errors.  Fig.  2  shows the H,H-K$_s$
colour magnitude diagram  (CMD) for our candidates as  well as for all
objects detected in the field. None of the candidates is characterised
by peculiar  colours with  respect to the  main sequence of  the field
stellar population, which suggest that they are main-sequence stars.

\begin{figure}
\centering 
\includegraphics[bb=30 150 570 700,height=8cm,angle=0,clip]{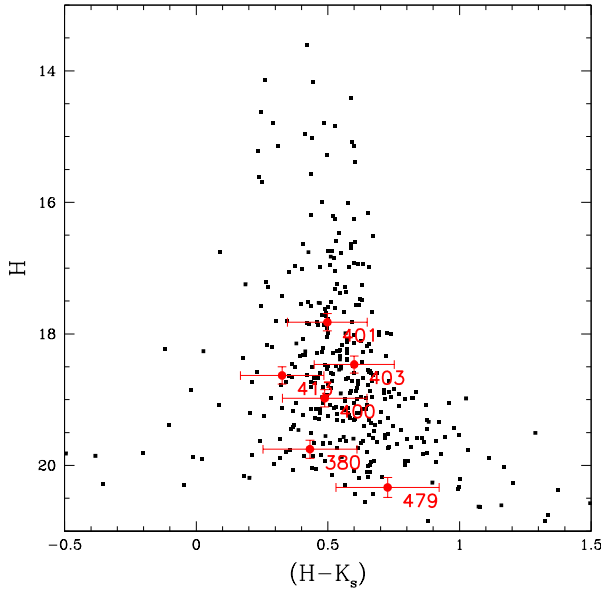}
  \caption{H,H-K$_s$ CMD of all stars detected in the \vlt$/$\naco\ field. All candidates identified in Fig. 1 are marked in red and labelled accordingly. No interstellar extinction correction has been applied.}
\label{rxj1713_cmd}       % Give a unique label
\end{figure}

\section{Discussion}

To determine whether  one of the detected objects is the
IR  counterpart  to 1WGA  J1713.4$-$3949,  we  investigated how  their
observed  properties fit  with different  scenarios.

\subsection{A binary system}

If our candidates are stars, we considered the possibility that one of
them is the companion of  the 1WGA J1713.4$-$3949 neutron star.  Their
observed colours are quite red (0.4$<$ H-K$_s$ $<$0.7), which suggests
that  they  might  be   intrinsically  red  late-type  stars.   To  be
compatible with the  observed range of H-K$_s$ (Ducati  et al.  2001),
e.g.  an M-type main sequence star  should be reddened by an amount of
interstellar  extinction  corresponding   to  an  $N_H  \sim  10^{22}$
cm$^{-2}$ (Predhel  \& Schmitt 1995).   This value is  compatible with
the  largest   values  obtained  from   the  spectral  fits   to  1WGA
J1713.4$-$3949 (Lazendic et al.   2003; Cassam-Chena\"i et al.  2004).
For  the originally  proposed 1WGA  J1713.4$-$3949 distance  of  6 kpc
(Slane  et al.   1999) an  M-type star  with such  an  high extinction
should  be at  least $\sim$0.7  magnitudes fainter  than  our faintest
candidate  (object  479).   An  early  to mid  M-type  star  would  be
compatible   with  the  revised   distance  of   $1.3  \pm   0.4$  kpc
(Cassam-Chena\"i  et  al.  2004)  but  it would  be  detected  in  our
\ntt/\susi\ image at I$\sim 20.2-21.7$.  Thus, we conclude that if our
candidates  are  stars  none  of  them can  be  associated  with  1WGA
J1713.4$-$3949.  Our  optical/IR magnitude upper limits  only allow an
undetected companion of spectral type later than M.

\subsection{An isolated neutron star}

If 1WGA J1713.4$-$3949 is indeed an  INS, we can then speculate if one
of our candidates  is the neutron star itself.  Due  to the paucity of
the neutron stars observed in the  IR (e.g. Mignani et al.  2007b) and
to the lack  of well-defined spectral templates, it  is very difficult
to estimate their expected IR  brightness. This is even more difficult
for  the CCO  neutron  stars,  none of  which  has been  unambiguosuly
identified so far (e.g. De Luca 2008).  In the best characterised case
of   rotation-powered  neutron   stars   one  can   deduce  that   the
magnetospheric  IR and  X-ray luminosities  correlate (Mignani  et al.
2007b; Possenti  et al.   2002).  By assuming,  e.g. a  blackbody plus
power  law  X-ray  spectrum   for  1WGA  J1713.4$-$3949  (Lazendic  et
al.   2003;  Cassam-Chena\"i  et   al.  2004)   we  then   scaled  the
magnetospheric IR-to-X-ray luminosity ratio  of the Vela pulsar, taken
as  a reference  because of  its  comparable age  ($\sim 10$  kyears).
After accounting for the corresponding interstellar extinction we thus
estimated K$_s  \sim 19.7$ for  1WGA J1713.4$-$3949, i.e.   similar to
the  magnitude  of object  479  (K$_s  \sim  19.6$).  Since  also  the
magnetospheric  optical   and  X-ray  luminosities   correlate  (e.g.,
Zharikov et al.  2004), we similarly estimated B$\approx28.3$ for 1WGA
J1713.4$-$3949 which,  however, is  below our \ntt/\susi\  upper limit
(B$\ge$26).   Thus, a neutron  star identification  can not  be firmly
excluded.

\subsection{A fossil disk}

As  discussed in Sect.1,  some CCO  models invoke  low-magnetised INSs
surrounded by  fallback disks.   So, the last  possibility is  that we
detected  the  IR  emission  from  such  a disk.   We  note  that  the
IR-to-X-ray  flux  ratio for  1WGA  J1713.4$-$3949  would be  $\approx
10^{-3}  - 10^{-2}$,  i.e.  much  larger than  that estimated  for the
anomalous X-ray pulsar  4U\,0142+61 (Wang et al.  2006),  the only INS
with evidence of  a fallback disk.  However, we can  not a priori rule
out  the  fallback  disk  scenario.   We computed  the  putative  disk
emission using  the model of Perna  et al. (2000),  which accounts for
both for the  contribution of viscous dissipation as  well as that due
to reprocessing of the neutron  star X-ray luminosity. As a reference,
we assumed  the X-ray luminosity  derived for the updated  distance of
$1.3 \pm 0.4$  kpc (Cassam-Chena\"i et al. 2004).   For a nominal disk
inclination angle of $60^\circ$ with respect to the line of sight, the
unknown model parameters  are the disk inner and  outer radii ($R_{\rm
in}$,$R_{\rm  out}$),   and  accretion  rate   ($\dot{M}$).   We  thus
iteratively  fitted   our  data  for  different  sets   of  the  model
parameters.  For  the dimmest  candidate we found  that the  IR fluxes
would  be consistent  with  a  spectrum of  a  disk ($R_{\rm  in}=0.28
R_\odot$, $R_{\rm out}=1.4R_\odot$) whose emission is dominated by the
reprocessed neutron  star X-ray luminosity (Fig. 3),  similarly to the
case of 4U\, 0142+61.  However, such  a disk should be detected in the
I  band, with a  flux $\sim  1.5$ magnitude  above our  measured upper
limit, as shown  in Fig. 3. The overprediction of  the optical flux is
even more dramatic for a  disk that fits the brighter counterparts. We
thus  conclude that,  if  the neutron  stra  has a  disk,  it was  not
detected by our observations.

\begin{figure}
\centering 
\includegraphics[bb=20 180 590 700,height=8cm,angle=0,clip]{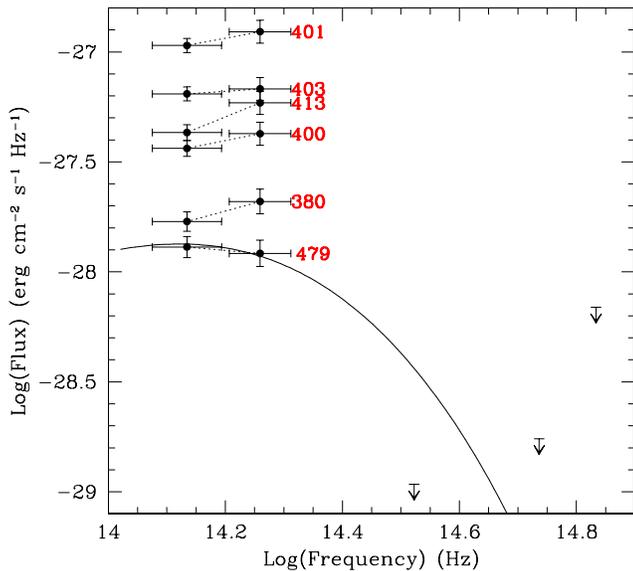}
  \caption{Dereddened IR spectra of the 1WGA J1713.4$-$3949 candidate counterparts. 
Dotted lines are drawn for guidance. The BVI bands upper limits are indicated. The solid line is the  best fitting disk spectrum 
($R_{\rm  in}=0.28 R_\odot$, $R_{\rm out}=1.4 R_\odot$) for object 479.  }
\label{rxj1713_cmd}       % Give a unique label
\end{figure}

\section{Conclusions}

We  performed  deep  optical  and  IR observations  of  the  CCO  1WGA
J1713.4$-$3949  in the G347.3-0.5  SNR, the  first ever  performed for
this source, with  the \ntt\ and the \vlt.  We  detected a few objects
close to the derived \chan\  X-ray error circle.  However, if they are
stars the  association with the CCO  would not be  compatible with its
current values of distance  and hydrogen column density.  Similarly to
the cases of the CCOs in  PKS 1209$-$51, Puppis A (Wang et al.  2007),
Cas A  (Fesen et al.  2006)  and RCW 103  (De Luca et al.   2008), our
results argue against  the presence of a companion  star, unless it is
later than M-type, and favour the INS scenario.  The identification of
the faintest candidate with the  neutron star itself can not be firmly
excluded, while the  identification with a fallback disk  is ruled out
by its non-detection  in the I band.  Thus, we  conclude that the 1WGA
J1713.4$-$3949 counterpart  is still unidentified.   Deeper optical/IR
observations  are  needed to  pinpoint  new  candidates. Although  the
source is apparently steady in  X-rays, flux variations as observed in
the RCW 103 CCO (Gotthelf et  al.  1999) can not be a priory excluded.
A prompt IR follow-up would  then increase the chances to identify the
1WGA J1713.4$-$3949 counterpart.

\begin{acknowledgements}

RPM  warmly thanks  N.  Ageorges  (ESO) for  her friendly  support  at  the  telescope, D.   Dobrzycka  (ESO)  for
reducing the IR data with the \naco\ pipeline.

\end{acknowledgements}

\end{document}